\documentclass[twocolumn,aps,pra,superscriptaddress,showpacs]{revtex4}
\usepackage{mathptmx}
\usepackage[T1]{fontenc}
\usepackage[latin9]{inputenc}
\usepackage{color}
\usepackage{verbatim}
\usepackage{amsmath}
\usepackage{amssymb}
\usepackage{graphicx}
\usepackage{esint}

\makeatletter
\@ifundefined{textcolor}{}
{%
 \definecolor{BLACK}{gray}{0}
 \definecolor{WHITE}{gray}{1}
 \definecolor{RED}{rgb}{1,0,0}
 \definecolor{GREEN}{rgb}{0,1,0}
 \definecolor{BLUE}{rgb}{0,0,1}
 \definecolor{CYAN}{cmyk}{1,0,0,0}
 \definecolor{MAGENTA}{cmyk}{0,1,0,0}
 \definecolor{YELLOW}{cmyk}{0,0,1,0}
}

\usepackage{siunitx}
\usepackage{bm}
\renewcommand{\vec}{\bm}

\makeatother

\begin{document}

\title{Emergence of metastable pointer states basis in non-Markovian quantum dynamics}

\author{F. Lastra}

\affiliation{Departamento de Física, Facultad de Ciencias Básicas, Universidad
de Antofagasta, Casilla 170, Antofagasta, Chile}

\author{C.E. L\'opez}

\affiliation{Departamento de Física, Universidad de Santiago de Chile, Universidad
de Santiago de Chile, Casilla 307, Correo 2, Santiago, Chile}

\affiliation{Center for the Development of Nanoscience and Nanotechnology, 9170124,
Estación Central, Santiago, Chile}

\author{S.A. Reyes}

\affiliation{Instituto de Física, Pontificia Universidad Católica de Chile, Casilla
306, Santiago 22, Chile}

\affiliation{Centro de Investigación en Nanotecnología y Materiales Avanzados
CIEN-UC, Pontificia Universidad Católica de Chile, Santiago, Chile}

\author{S. Wallentowitz}

\affiliation{Instituto de Física, Pontificia Universidad Católica de Chile, Casilla
306, Santiago 22, Chile}

\affiliation{Centro de Investigación en Nanotecnología y Materiales Avanzados
CIEN-UC, Pontificia Universidad Católica de Chile, Santiago, Chile}
\begin{abstract}
We investigate the dynamics of classical and quantum correlations
between two qubits. Each qubit is implemented by a pair of phosphorous
impurities embedded in a silicon substrate. The main decoherence mechanism
affecting these types of qubits is provided by the coupling of the
phosphorous impurities to the acoustical vibrations of the silicon
lattice. We find that depending on the temperature of the substrate
and the initial state, three different dynamics can be found. These
are characterized by the number of abrupt changes in both classical
and quantum correlations. We also show that the correlations do not
disappear. Moreover, before the classical correlations reach a constant
value, they may experience successive abrupt changes associated with
the apparition of metastable pointer states basis. Then, a constant value for the
classical correlations is reached when the preferred basis is established.
\end{abstract}
\maketitle
Recently the study of classical and quantum correlations has become
a central subject of investigation. The study of quantum correlations
between quantum systems is a problem as old as the quantum theory.
For many years, it was widely believed that quantum entanglement would be the only relevant type of correlation for quantum information protocols.
However, it has been proved that efficient quantum protocols can be
performed in the absence of entanglement \cite{Ent}. A bipartite
quantum system A-B can feature both quantum and classical correlations
between its constituent parts A and B, respectively. All these correlations
can be characterized by the quantum mutual information \cite{Ollivier01,Gr}

\begin{equation}
I(\hat{\rho}_{{\rm AB}})=S(\hat{\rho}_{{\rm A}})+S(\hat{\rho}_{{\rm B}})-S(\hat{\rho}_{{\rm AB}}),\label{qmi}
\end{equation}
where $S(\hat{\rho})=-{\rm Tr}[\hat{\rho}\lg(\hat{\rho})]$ is the
von Neumann entropy \cite{Nielsen2000}. Based on this expression
it is commonly believed\textcolor{blue}{{} }that the correlations can
be\textcolor{blue}{{} }separated according to their classical and quantum
nature, respectively \cite{Ollivier01}. In this way the quantum discord
can be introduced as \cite{Ollivier01,Gr} and  \cite{Zurek00,Henderson01,Opp,Luo}

\begin{equation}
D(\hat{\rho}_{AB})=I(\hat{\rho}_{AB})-C(\hat{\rho}_{AB})\label{qd}
\end{equation}
where $C(\hat{\rho}_{AB})$ are the classical correlations \cite{Ollivier01}, \cite{Henderson01} defined
by the following maximization procedure: A complete set of projector
operators $\{\hat{\Pi}_{k}\}$ must be constructed for the subsystem
B. Then the quantity 
\begin{equation}
C(\hat{\rho}_{AB})=\max_{\{\hat{\Pi}_{k}\}}\left[S(\hat{\rho}_{A})-S(\hat{\rho}_{AB}\mid\{\hat{\Pi}_{k}\})\right],\label{cc}
\end{equation}
must be maximized with respect to variation of the set of $\{\hat{\Pi}_{k}\}$
where $S(\hat{\rho}_{AB}\mid\{\hat{\Pi}_{k}\})=\sum_{k}p_{k}S(\hat{\rho}_{k})$,
$p_{k}={\rm Tr}(\hat{\rho}_{AB}\hat{\Pi}_{k})$, and $\hat{\rho}_{k}={\rm Tr}_{B}(\hat{\Pi}_{k}\hat{\rho}_{AB}\hat{\Pi}_{k})/p_{k}$.

The competition between classical and quantum correlations
parts of a quantum system has attracted increasing attention in recent
years. It is widely believed that for decoherence, caused by contact
with an external reservoir, both types of correlations of a system
are continuously depleted and transferred to the reservoir degrees
of freedom. This depletion happens on a characteristic timescale that
defines the decoherence time of the system, which is the effective
feature of the system's coupling to the reservoir. An alternative
way to characterize the decoherence can be obtained from the dynamical
raise of the system's entropy.

However, it has been pointed out that the former case is not the general
one, as dynamical transitions from quantum to classical correlations
may also be discontinuous \cite{Maz,Mazz}. This has been demonstrated
for a system experiencing decoherence by a dephasing channel. Furthermore,
it has been shown experimentally \cite{Xu2010,Cor,Pau}, that classical
correlations may change abruptly to a non-vanishing stationary value.
This feature has been proposed to determine whether a quantum system
has reached its classical regime or not \cite{Cor}. The appearance
of a stationary classical correlation is associated by the emergence
of a basis of pointer states in one of the two subsystems \cite{Cor},
denoted also as ``preferred basis for the apparatus'', where the
apparatus may be understood as one of the two subsystems \cite{Zurek1}.

In this manuscript, we show analytically that not only the quantum-to-classical
transition may be discontinuous, but that these abrupt changes may
also occur various times during the dynamical evolution of the system.
Therefore, there may exist intermediate stages with constant classical
correlation. Each stage defines a distinct and metastable basis of
pointer states. Only after passing through these stages --- after
a sufficiently long time --- the final basis of pointer states is
asymptotically established. 

Our system is composed of two charge quantum bits (qubits), implemented by electrons localized at donor impurities, i.e. P, that are embedded in a semiconductor host, i.e. Si \cite{Hol}. Each qubit consists of a pair of
impurities sharing a single electron. The quantum information is encoded
as the position of the electron within the pair of impurities. At
or below room temperature the dominant source of decoherence of this
implementation of qubits is off-resonant scattering of acoustical
phonons in the substrate. This decoherence mechanism has been shown,
both numerically \cite{Eckel06} and in analytically closed form \cite{Lastra11a,Lastra11b},
to provide a non-Markovian dephasing of the qubits. It also induces
a disentanglement of pairs of qubits with subsequent partial recovering
of the initial entanglement \cite{Lastra12}.

At sufficiently large distances between the qubits, Coulomb repulsion
and cross tunneling of electrons can be neglected. Then the dynamics
is generated by off-resonant scattering of acoustical phonons at the
donor-based charge qubits, which is described by the spin-boson Hamiltonian,
\begin{eqnarray}
H & = & \hbar\sum_{b}(\omega_{b}\hat{S}_{b,z}+\Delta_{b}\hat{S}_{b,{\color{blue}{\normalcolor x}}})+\sum_{\vec{k}}\hbar\nu_{k}\hat{a}_{\vec{k}}^{\dagger}\hat{a}_{\vec{k}}\nonumber \\
 &  & +\hbar\sum_{b}\sum_{\vec{k}}\hat{S}_{b,z}(g_{b,\vec{k}}\hat{a}_{\vec{k}}^{\dagger}+g_{b,\vec{k}}^{*}\hat{a}_{\vec{k}}).\label{ham}
\end{eqnarray}
Here the spin-$\frac{1}{2}$ operators $\hat{\vec{S}}_{b}$ act on
the states $|m_{b}\rangle$ of the $b$th qubit, where $m_{b}=\pm\frac{1}{2}$
denotes the localization of the qubit's electron at one of the two
impurity sites. Furthermore, $\hat{a}_{\vec{k}}$ are the bosonic
annihilation operators of longitudinal acoustical phonons of wave
vector $\vec{k}$ and linear dispersion relation $\nu_{k}=sk$ with
$s$ being the speed of sound in the substrate. The qubits have transition
frequencies $\omega_{b}$ and tunneling rates $\Delta_{b}$, and are
interacting with the phonons via the coupling rate 
\[
g_{b,\vec{k}}=\frac{D}{\hbar s}\sqrt{\frac{2\hbar\nu_{k}}{M_{o}}}\sum_{m_{b}=\pm\frac{1}{2}}\frac{m_{b}e^{-i\vec{k}\bullet(\vec{r}_{b}+m_{b}\vec{d}_{b})}}{\left[1+\left(\frac{ka_{{\rm B}}}{2}\right)^{2}\right]^{2}}.
\]
where $D$ is the deformation constant of the substrate, $M_{0}$
is the unit cell mass, and $a_{{\rm B}}$ is the corresponding Bohr
radius in the substrate.

\begin{figure}
\begin{centering}
\includegraphics[width=0.8\columnwidth]{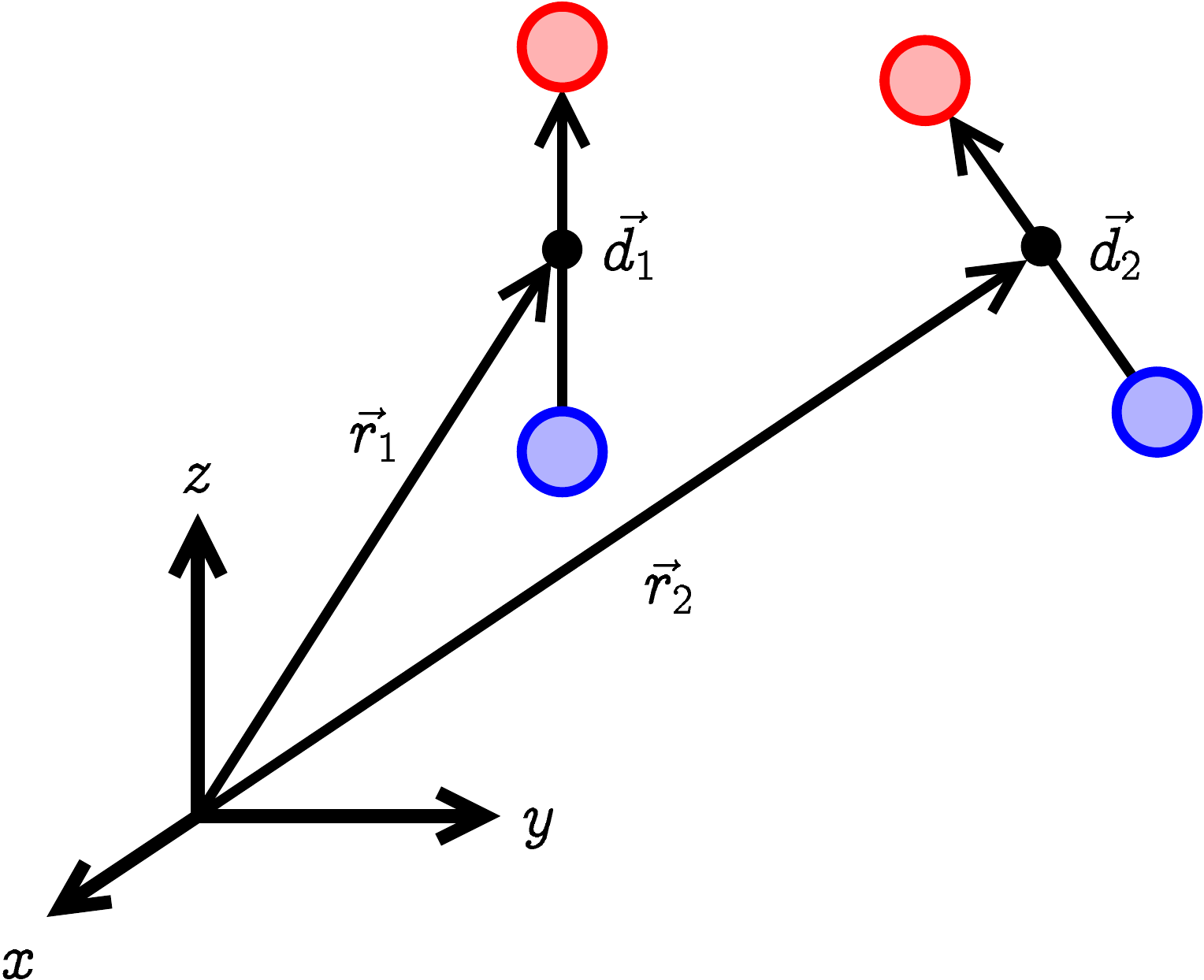}
\par\end{centering}

\caption{\label{fig:Schematic-outline...}Schematic outline of the two donor-based
charge qubits ($b=1,2$) formed by 4 donor sites in a semiconductor material.
Blue and red donor sites correspond to qubit states $m_{b}=-\frac{1}{2}$
and $m_{b}=+\frac{1}{2}$, respectively. Each qubit $b$ is located
at position $\vec{r}_{b}$ and has an inter-site distance vector $\vec{d}_{b}$
pointing from site $m_{b}=-\frac{1}{2}$ to site $m_{b}=+\frac{1}{2}$.}

\end{figure}

The geometrical configuration of the N qubit setup is characterized
by a vector $\hat{r}_{b}$ with $b=1,2,3,\ldots$ which labels the
position of the center of each qubit, $\hat{d}_{b}$ is the inter
donor distance, for the sake of simplicity, this parameter as well
as Bohr radius are taken to be the same for each qubit, see Figure
\ref{fig:Schematic-outline...}. Here also, we assume the distance
between two adjacent qubit is larger than the inter donor distance,
this provides a condition for preventing inter qubit tunneling.

Applying electric potentials via suitably positioned electrodes, the
qubit's transition frequencies can be raised to surpass the tunneling
rates, $\omega_{b}\gg\Delta_{b}$, and tunneling can be neglected
so that fully analytic solutions of the spin-boson dynamics can be
found. Assuming then that the thermal excitation of substrate phonons
is sufficiently low to not excite the qubit transitions, $k_{{\rm B}}T\ll\hbar\omega_{b}$,
each qubit can be individually manipulated to prepare a general initial
state of the qubits factorized with respect to the phonon state,

\begin{figure*}
\centering{}\includegraphics[clip,width=0.33\textwidth]{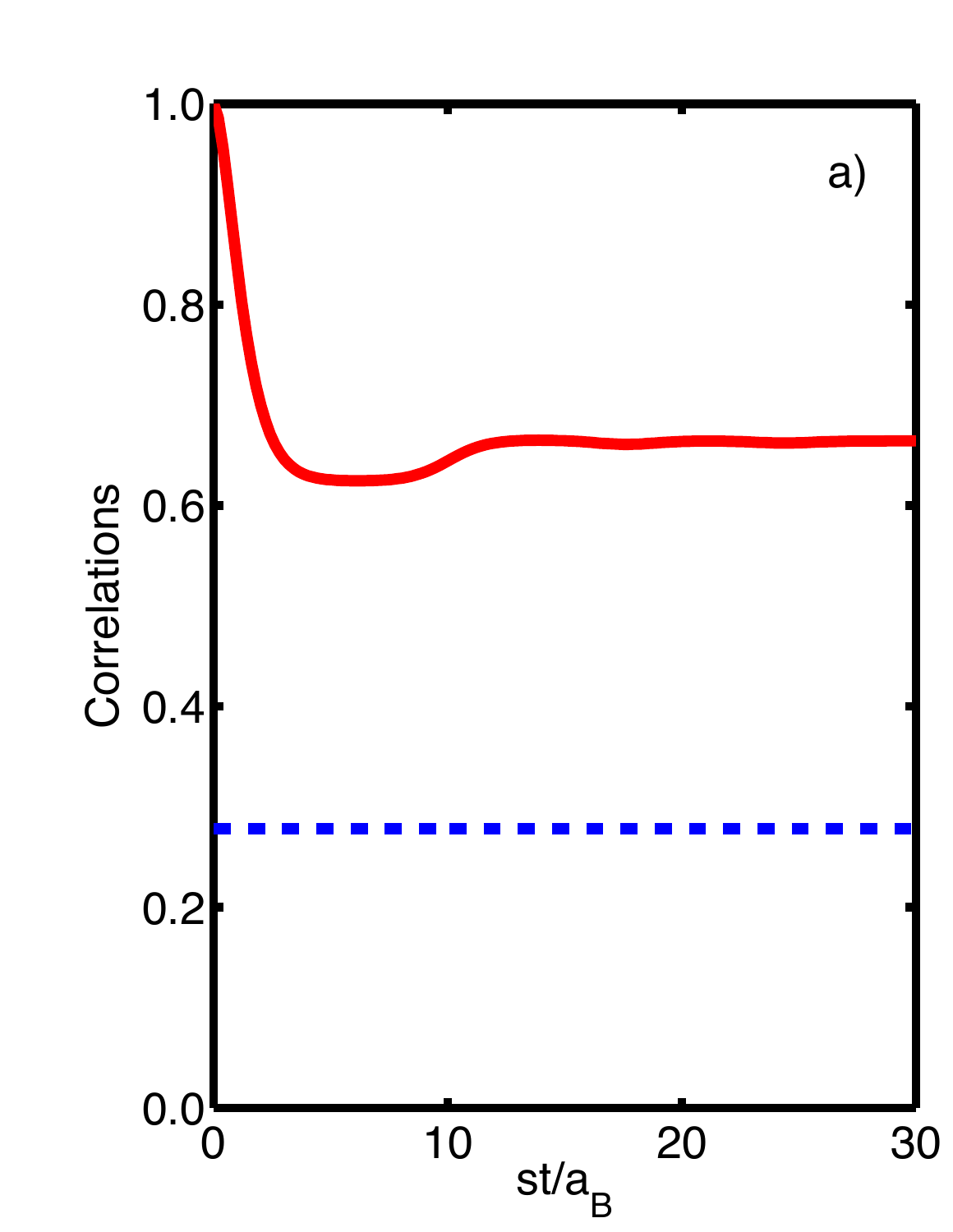}
\includegraphics[clip,width=0.33\textwidth]{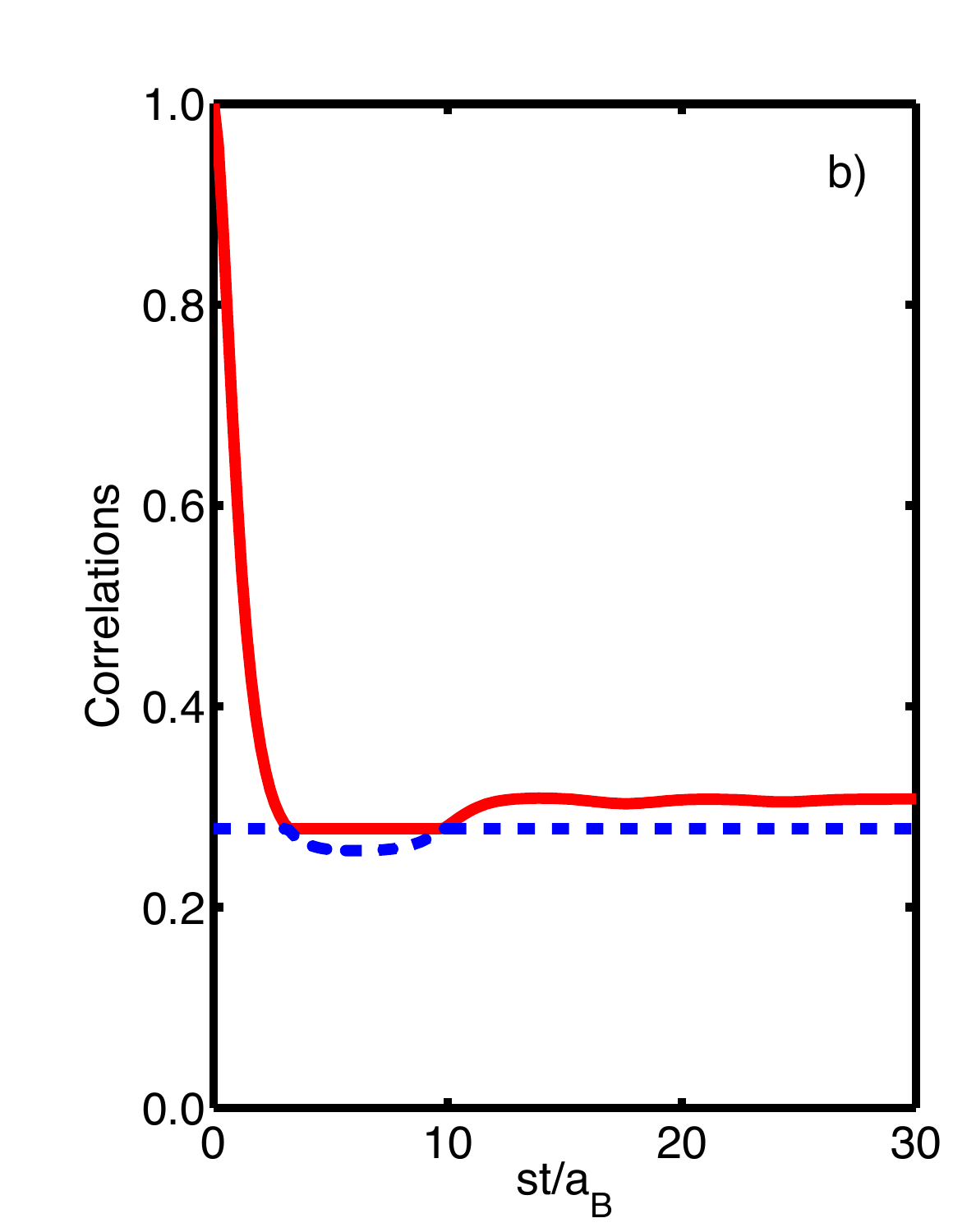} \includegraphics[clip,width=0.33\textwidth]{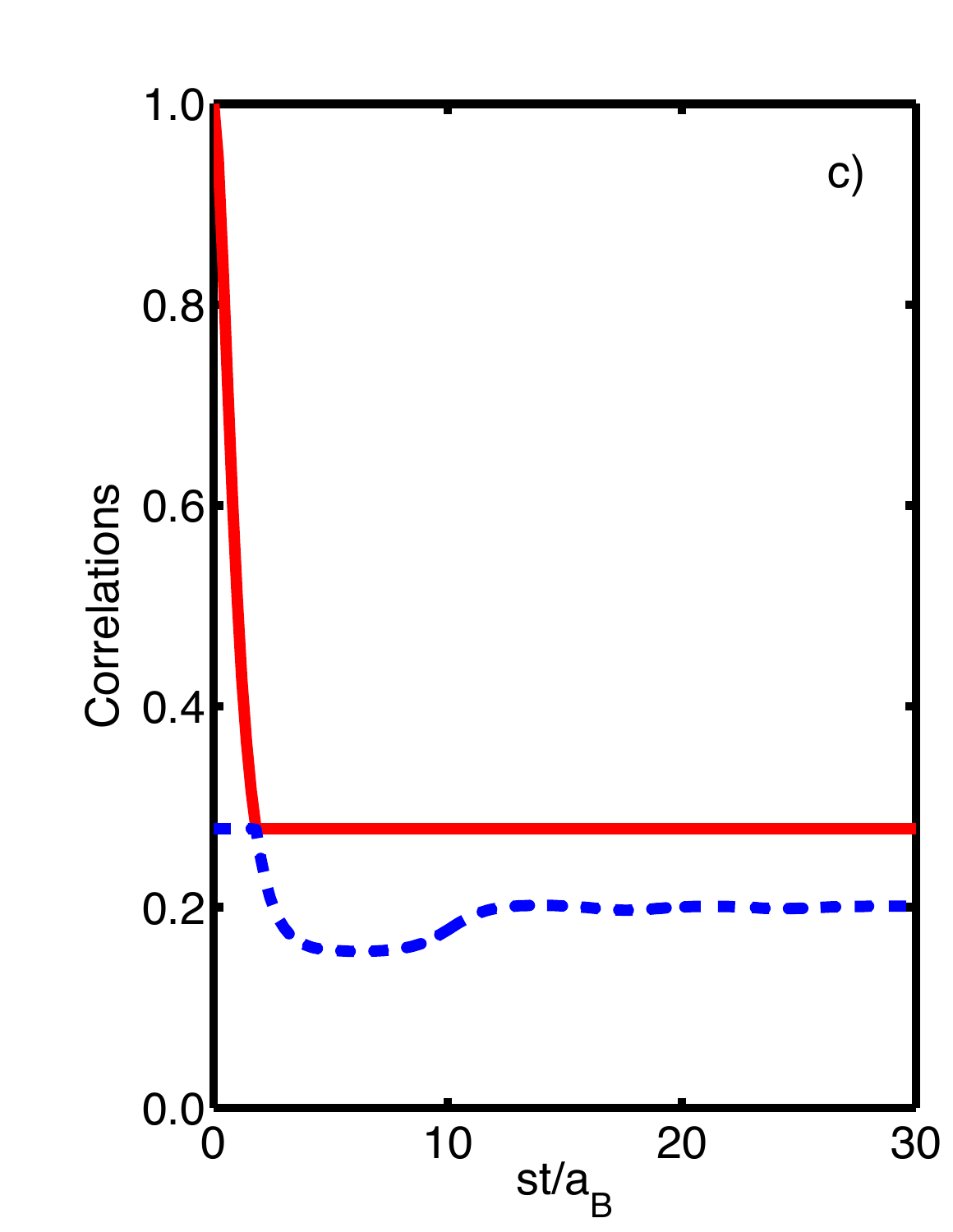}
\caption{\label{fig:Evolution-of-the}Evolution of the classical correlation
(red solid-line), the quantum discord (blue dotted-line) for an initial
state characterized by $p=0.8$ and different temperatures, a) $T/T_{B}=0.01$,
b) $T/T_{B}=0.035$, c) $T/T_{B}=0.05$. Geometrical parameters are:
$|\vec{d}_{1}|=|\vec{d}_{2}|=10a_{{\rm B}}$, $|\vec{r}_{1}-\vec{r}_{2}|=20a_{{\rm B}}$,
$\angle(\vec{d}_{1},\vec{d}_{2})=\SI{45}{\degree}$.}
\end{figure*}

\begin{equation}
\hat{\varrho}_{{\rm ph+qubits}}(0)=\sum_{\{m_{b}\},\{s_{b}\}}\rho_{\{m_{b}\},\{s_{b}\}}(0)|\{m_{b}\}\rangle\langle\{s_{b}\}|\otimes\hat{\varrho}_{{\rm ph,th}},\label{eq:initial-state}
\end{equation}
where $|\{m_{b}\}\rangle=|m_{1}\rangle\otimes|m_{2}\rangle$ are the
bipartite qubit states and $\hat{\varrho}_{{\rm ph,th}}$ is the thermal
state of the phonons. The reduced density matrix of the qubits can
then be shown to evolve in time as \cite{Lastra11a,Lastra11b,Lastra12}

\begin{equation}
\rho_{\{m_{b}\},\{s_{b}\}}(t)=\rho_{\{m_{b}\},\{s_{b}\}}(0)f_{\{m_{b}\},\{s_{b}\}}(t).\label{eq:rho-offdiag2}
\end{equation}
where the time dependence is given by the functions \cite{Lastra12}

\begin{eqnarray}
f_{\{m_{b}\},\{s_{b}\}}(t)=\exp\left[-\int_{0}^{t}dt^{\prime}\Gamma_{\{m_{b}\},\{s_{b}\}}(t^{\prime})\right].\label{eq:efe}
\end{eqnarray}
Here the decoherence rate of the bipartite qubit state results as
\begin{eqnarray}
\Gamma_{\{m_{b}\},\{s_{b}\}}(t) & = & \sum_{b,b'}(m_{b}-s_{b})(m_{b'}-s_{b'})\gamma_{b,b'}(t),\label{eq:decrate}
\end{eqnarray}
where the inter-qubit decorrelation rate is given by 
\begin{eqnarray}
\gamma_{b,b'}(t) & =4 & \sum_{m_{b}}\sum_{s_{b^{\prime}}}m_{b}s_{b^{\prime}}\gamma\left(t;|(\vec{r}_{b}+m_{b}\vec{d}_{b})-(\vec{r}_{b^{\prime}}+s_{b^{\prime}}\vec{d}_{b^{\prime}})|\right).\label{eq:inter-bit-rate}
\end{eqnarray}
The inter-donor decoherence rate is defined as 
\[
\gamma(t;l)=\frac{2\pi s}{l}\frac{T}{T_{{\rm s}}}\sum_{\sigma=\pm1}\sigma\left(\frac{x^{3}}{6}+\frac{x^{2}}{2}+\frac{5x}{8}+\frac{5}{16}\right)e^{-2x},
\]
with $x=|l-\sigma st|/a$. This decoherence rate is proportional to
the substrate temperature $T$, where the temperature scale is defined
in terms of parameters of the substrate as $T_{{\rm s}}=a\rho_{{\rm s}}s^{4}h^{2}/(k_{{\rm B}}D^{2})$,
with $\rho_{{\rm s}}$ being the substrate mass density.

As the initial qubit state we consider a statistical mixture of two
Bell states ($p\in[0,1]$), 
\begin{equation}
\hat{\rho}_{{\rm qubits}}(0)=p\mid\Psi_{+}\rangle\langle\Psi_{+}\mid+(1-p)\mid\Phi_{+}\rangle\langle\Phi_{+}\mid.\label{eq:qubit-state}
\end{equation}
Thus, the initial state can be expressed in the basis \{$|\frac{1}{2},\frac{1}{2}\rangle$,
$|\frac{1}{2},-\frac{1}{2}\rangle$, $|-\frac{1}{2},\frac{1}{2}\rangle$,
$|-\frac{1}{2},-\frac{1}{2}\rangle$\} as an $X$ state and remains in
this form throughout the non-dissipative time evolution, as 
\[
\rho_{{\rm qubits}}(t)=\frac{1}{2}\begin{bmatrix}p & 0 & 0 & b(t)\\
0 & 1-p & c(t) & 0\\
0 & c(t) & 1-p & 0\\
b(t) & 0 & 0 & p
\end{bmatrix}
\]
where we defined $b(t)=pf_{\frac{1}{2},\frac{1}{2};-\frac{1}{2},-\frac{1}{2}}(t)$
and $c(t)=(1-p)f_{\frac{1}{2},-\frac{1}{2};-\frac{1}{2},\frac{1}{2}}(t)$.
This density matrix reaches a stationary state for $t\gg d/s$, where
$d$ is the characteristic distance between donor sites, and $b(t)$
and $c(t)$ attain constant values that depend on the geometric configuration
of the qubit pair. In general analytical solutions for the correlations
of a two-qubit state are yet unknown. However, for the X state present
here, analytical solutions for the correlations can be found following
the lines of Ref. \cite{xstate}. The result for the classical correlations
is
\begin{equation}
C(t)=1-K\left(w(t)\right),\label{eq:C-analytical}
\end{equation}
where 
\begin{equation}
w(t)=\max\left[|a|,b+c\right].\label{eq:w-def}
\end{equation}
with $a=2p-1$ and the function $K$ reads 
\[
K(x)=-\frac{1+x}{2}\lg\left(\frac{1+x}{2}\right)-\frac{1-x}{2}\lg\left(\frac{1-x}{2}\right).
\]
Furthermore, the quantum discord is obtained as %
\begin{comment}
\begin{eqnarray}
D(t) & = & C(t)+p\lg(p)+(1-p)\lg(1-p)\nonumber \\
 &  & +pK\left(\frac{c(t)}{1-p}\right)+(1-p)K\left(\frac{b(t)}{p}\right).\label{eq:D-analytical}
\end{eqnarray}
\end{comment}
\begin{eqnarray}
D(t) & = & 1+p\lg(p)+(1-p)\lg(1-p)+K(w(t))\nonumber \\
 &  & -pK\left(\frac{b(t)}{p}\right)-(1-p)K\left(\frac{c(t)}{1-p}\right).\label{eq:D-analytical-1}
\end{eqnarray}

The classical correlation (\ref{eq:C-analytical}) and the quantum
discord (\ref{eq:D-analytical-1}) of the state (\ref{eq:qubit-state})
are shown in Fig. \ref{fig:Evolution-of-the} as functions of time.
It can be seen that three cases exist depending on the temperature.
For very low temperature the classical correlation is well above the
quantum discord, the latter being constant, see Fig. \ref{fig:Evolution-of-the}
a). Increasing slightly temperature, the minimum of the classical
correlation merges with quantum discord, so that the latter now shows
a minimum, as seen in Fig. \ref{fig:Evolution-of-the} b). Further
increasing the temperature leads to a third case where the classical
correlation becomes constant after an initial decay, see Fig. \ref{fig:Evolution-of-the}
c). The latter case has been observed previously for dephasing qubits
\cite{Maz,Mazz,Xu2010,Cor}.

Whereas the above results have been obtained in a general way, we
may also derive these results by explicitly performing the variation
over the set of projectors in Eq. (\ref{cc}). In this way a basis
of two orthogonal states is obtained that maximizes the classical
information at a certain instant of time. Whenever the classical information
is constant as a function of time, this basis is the
basis of pointer states. To study these features in more detail, we
may analytically determine the basis of pointer states, when it exists.
Without loss of generality, the subsystem B is considered as an apparatus
that performs measurements on subsystem A. Defining for subsystem
B the arbitrary basis of two orthonormal states, $|\psi_{1}\rangle_{B}=\cos(\theta/2)|{\textstyle -\frac{1}{2}}\rangle_{B}+e^{i\phi}\sin(\theta/2)|{\textstyle \frac{1}{2}}\rangle_{B}$
and $|\psi_{2}\rangle_{B}=\cos(\theta/2)|{\textstyle \frac{1}{2}}\rangle_{B}-e^{-i\phi}\sin(\theta/2)|{\textstyle -\frac{1}{2}}\rangle_{B}$,
the measurement projectors can be constructed as $\hat{\Pi}_{k}=|\psi_{k}\rangle_{BB}\langle\psi_{k}|$
($k=1,2$).

The classical information is of the form 
\begin{equation*}
C(t)=\max_{\{\theta,\phi\}} [G(\theta,\phi,t)],
\end{equation*}
where 
\begin{equation*}
G(\theta,\phi,t)=1+\frac{1}{2}\left[(1+g)\lg(1+g)+(1-g)\lg(1-g)\right].
\end{equation*}
with
\begin{equation*}
g(\theta,\phi,t)=\sqrt{a^{2}\cos^{2}\theta+\sin^{2}\theta\left[b^{2}+c^{2}+2bc\cos(2\phi)\right]},
\end{equation*}
As a consequence it can be shown that the extrema of $G(\theta,\phi,t)$
with respect to the position on the Bloch sphere ($\theta,\phi$)
coincide with the corresponding extrema of $g(\theta,\phi,t)$. Note,
that since $b,c\geq0$, $g$ attains its possible maxima only for
$\phi=0,\pi$. Thus, maximizing $g$ is equivalent to maximizing the
expression 
\[
a^{2}\cos^{2}\theta+\left[b(t)+c(t)\right]^{2}\sin^{2}\theta.
\]

For time intervals where $a>b(t)+c(t)$ the maxima of the classical
correlation appears at the poles of the Bloch sphere, $\theta=0,\pi$.
In this time intervals, the classical information is constant, as
seen from Eqs (\ref{eq:C-analytical}) and (\ref{eq:w-def}). Therefore,
this basis of eigenstates of $\hat{\sigma}_{z}$ corresponds to a
basis of pointer states.

On the other hand, if $a<b(t)+c(t)$ the maxima appear at $\theta=\frac{\pi}{2}$
so that the basis corresponds to eigenstates of $\hat{\sigma}_{x}$.
However, the classical information in general is not constant. Only
for very large times, when $b(t)$ and $c(t)$ reach their stationary
values, the classical information asymptotically reaches a constant
value. This latter case again corresponds to a basis of pointer states.
As $b(t)$ and $c(t)$ depend on time via the functions (\ref{eq:efe})
and since these functions scale with temperature, varying the temperature
allows a change of the basis of pointer states. This can be seen in
the three cases depicted in Fig. \ref{fig:Evolution-of-the}: In a)
for large times the classical correlation becomes stationary and a
basis of pointer states is asymptotically established as eigenstates
of $\hat{\sigma}_{x}$. Increasing temperature, rf. b), a metastable
regime of $\hat{\sigma}_{z}$ pointer states is observed for the time
interval where $C$ is constant, which is replaced for large times
by the $\hat{\sigma}_{x}$ pointer state basis. Finally, for even
higher temperatures the previously metastable regime becomes stable
and the $\hat{\sigma}_{z}$ eigenstates become the pointer states,
see c).

\begin{figure}[t]
\centering{}
\includegraphics[width=1\columnwidth]{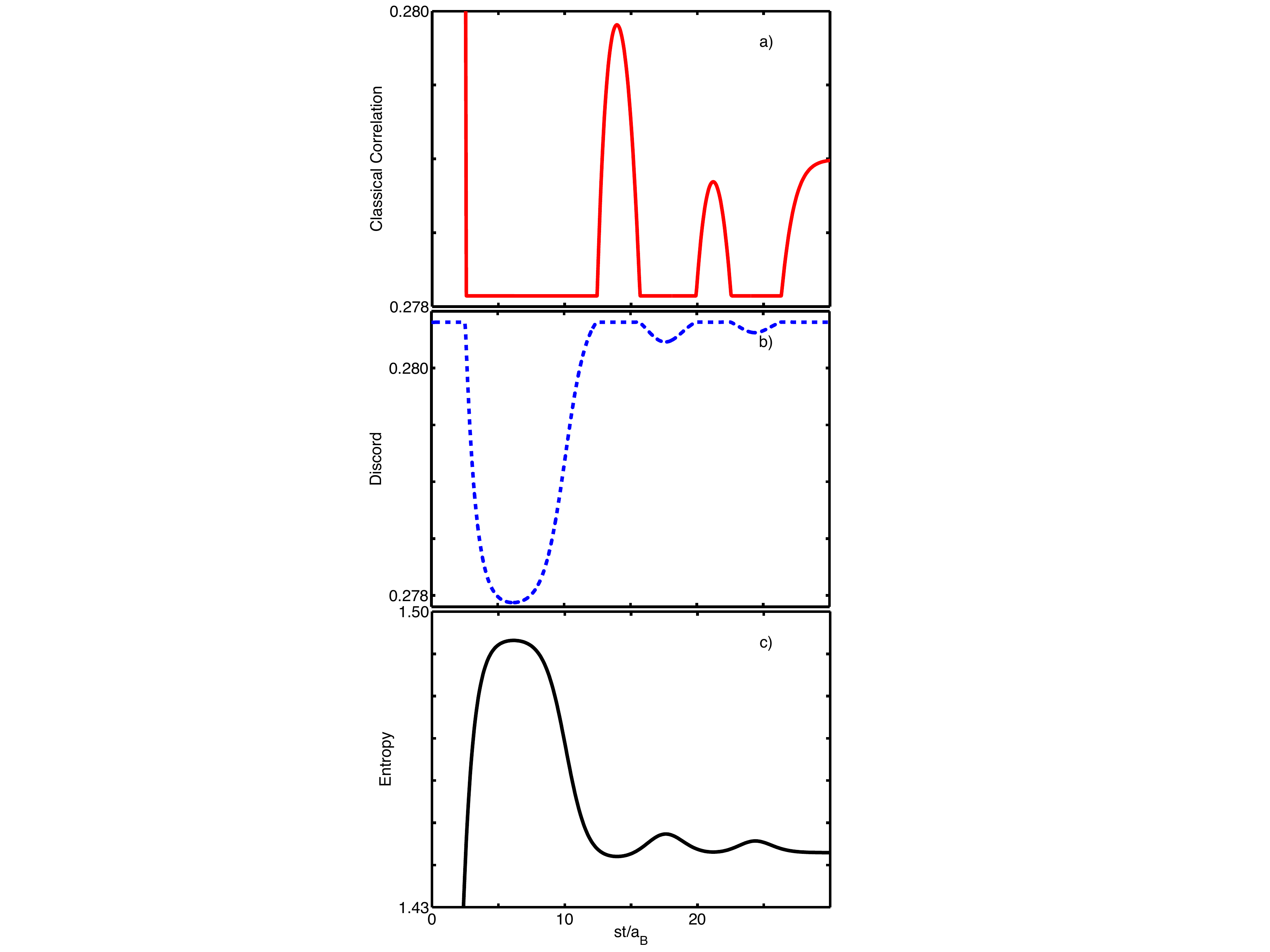}
\caption{\label{fig:Evolution-of-the-1}Evolution of the classical correlation
(red solid-line), the quantum discord (blue dashed-line) and Von Neumann
entropy (black solid-line) for an initial state with mixing parameter
$p=0.8$ and $T/T_{B}=0.0384$.}
\end{figure}

At intermediate temperatures between cases b) and c), i.e. when the
metastable $\hat{\sigma}_{z}$ pointer state basis gradually becomes
stable, an interesting feature can be observed, as shown in Fig. \ref{fig:Evolution-of-the-1}.
When increasing temperature more than one metastable regime can occur,
e.g. three regimes in the case shown in Fig. \ref{fig:Evolution-of-the-1}.
For long times again a stationary $\hat{\sigma}_{x}$ pointer-state
basis is reached asymptotically. The appearance of one or more metastable
regimes is due to the non-Markovian nature of the evolution of the
system \cite{Lastra11a,Lastra11b,Lastra12}. Indeed, we have checked that increasing the separation between the qubits, the quantum correlations decay monotonically before reaching a stationary (non zero) value. This dynamics resembles the one in Markovian systems. Also, the metastable pointer states do not appear in this regime. The reason is the following: the probability of a photon emitted by a qubit is absorbed by the other qubit, decreases when increasing separation between qubits. In other words, the phonons are more likely to spread around the lattice, leading to a Markovian dynamics. Related to the work of authors in ref.  \cite{Pau} where double sudden transitions in geometric quantum correlations are reported, in our case we found multiple sudden transitions for quantum discord (see Fig. 3 b)). 

The order of magnitude of the temperature $T_{{\rm P}}$ where the abrupt transition between
$\hat{\sigma}_{z}$ and $\hat{\sigma}_{x}$ pointer states occur, can
be estimated from the order of magnitude of the stationary values
$b(\infty)\approx p\exp(-16\pi T/T_{{\rm s}})$ and $c(\infty)=(1-p)\exp(-16\pi T/T_{{\rm s}})$,
and results from the condition $|a|=b(\infty)+c(\infty)$ as 
\begin{equation}
T_{{\rm P}}/T_{{\rm s}}\approx-\frac{\ln\left|2p-1\right|}{16\pi}.\label{eq:TP}
\end{equation}
Indeed this estimate agrees with the fact that for the special case
of a pure initial state ($p=1/2$) no abrupt transition can occur. 

The appearance of metastable pointer state basis is also evidenced
by the time evolution of the Von Neumann entropy $S(\hat{\rho}_{{\rm AB}})$.
Within each metastable regime the Von Neumann entropy reaches a local
maximum, which is in agreement with the meta stability of the pointer-state
basis. At large times the entropy, similar to the classical correlation,
reaches a stationary value.

We have shown that for a system of two donor-based charge qubits,
the decoherence may lead to a time evolution where a series of transitory
stages appear, each stage establishing a characteristic basis of pointer
states. This scenario is fundamentally different from single abrupt
transitions of the classical correlation, as shown in Refs. \cite{Maz,Mazz}.
However, the latter case can be observed in our system for the case
of higher temperatures, as shown in Fig. \ref{fig:Evolution-of-the}
c). The system described here, exhibits a phase-like transition with respect
to a change of temperature, where the equilibrium basis of pointer
states go from $\hat{\sigma}_{x}$ eigenstates at $T\ll T_{{\rm P}}$
to $\hat{\sigma}_{z}$ eigenstates at $T\gg T_{{\rm P}}$, where $T_{{\rm P}}$
is estimated in Eq. (\ref{eq:TP}). This feature opens the possibility
to engineer the basis of pointer states by tuning the parameters of
the physical system.

FL acknowledges support by Fondecyt 11110277, CEL by Fondecyt 1121034,
PBCT-CONICYT PSD54 and Financiamiento Basal para Centros Científicos
y Tecnológicos de Excelencia, SAR by FONDECYT 11110537, and SW by
PUC PUENTE 27/2013.


\begin{thebibliography}{10}
\bibitem{Ent}B.P. Lanyon, M. Barbieri, M.P. Almeida and A.G. White,
Phys. Rev. Lett. \textbf{101}, 200501 (2008).

\bibitem{Ollivier01}H. Ollivier and W.H. Zurek, Phys. Rev. Lett.
\textbf{88}, 017901 (2001).

\bibitem{Gr}B. Groisman, S. Popescu and A. Winter, Phys. Rev. A
\textbf{72} 032317 (2005).

\bibitem{Nielsen2000}M.A. Nielsen and I.L. Chuang, \emph{Quantum
Computation and Quantum Information} (Cambridge University Press,
Cambridge, 2000).

\bibitem{Zurek00}W.H. Zurek, Annalen der Physik \textbf{9}, 855 (2000).

\bibitem{Henderson01}L. Henderson and V. Vedral, J. Phys. A \textbf{34},
6899 (2001).

\bibitem{Opp}J. Oppenheim, M. Horodecki, P. Horodecki and R. Horodecki, Phys. Rev. Lett. \textbf{89}, 180402
(2002).

\bibitem{Luo}S. Luo, Phys. Rev. A \textbf{77}, 042303 (2008).

\bibitem{Maz}J. Maziero, L.C. Celeri, R.M. Serra and V. Vedral,
Phys. Rev. A \textbf{80}, 044102 (2009).

\bibitem{Mazz}L. Mazzola, J. Piilo and S. Maniscalco, Phys. Rev.
Lett. \textbf{104}, 200401 (2010).

\bibitem{Xu2010}J.S. Xu, C.F. Li, C.J. Zhang, X.Y. Xu, Y.S. Zhang
and G.C. Guo, Phys. Rev. A \textbf{82}, 042328 (2010).

\bibitem{Cor}M.F. Cornelio, O. J. Farias, F.F. Fanchini, I. Frerot,
G.H. Aguilar, M.O. Hor-Meyll, M.C. de Oliveira, S.P. Walborn, A.O.
Caldeira and P. H. Souto Ribeiro, Phys. Rev. Lett. \textbf{109}, 190402
(2012).

\bibitem{Pau}F. M. Paula, I. A. Silva, J. D. Montealegre, A. M. Souza, E. R. deAzevedo, R. S. Sarthour, A. Saguia, I. S. Oliveira, D. O. Soares-Pinto, G. Adesso and M. S. Sarandy, Phys. Rev. Lett. \textbf{111}, 250401
(2013).

\bibitem{Zurek1}W.H. Zurek, Phys. Rev. D \textbf{24}, 1516-1525 (1981).

\bibitem{Hol}L. C. L. Hollenberg, A. S. Dzurak, C. Wellard, A. R. Hamilton, D.J. Reilly, G. J. Milburn and R. G. Clark, Phys. Rev. B \textbf{69}, 113301 (2004).

\bibitem{Eckel06}J. Eckel, S. Weiss and M. Thorwart, Eur. Phys.
J. B \textbf{53}, 91 (2006).

\bibitem{Lastra11a}F. Lastra, S.A. Reyes and S. Wallentowitz, J.
Phys. B \textbf{44}, 015504 (2011).

\bibitem{Lastra11b}F. Lastra, S.A. Reyes and S. Wallentowitz, Rev.
Mex. Fis. S \textbf{57}, 148 (2011).

\bibitem{Lastra12}F. Lastra, S.A. Reyes and S. Wallentowitz, J.
Phys. B \textbf{45}, 015503 (2012).

\bibitem{xstate}M. Ali, A.R.P. Rau and G. Alber, Phys. Rev. A \textbf{81},
042105 (2010); \emph{ibid.} \textbf{82}, 069902(E) (2010).
\end{thebibliography}
\end{document}